\documentclass{article}

\PassOptionsToPackage{numbers,compress}{natbib}
\usepackage[nonanonymous]{neurips_2026}
\makeatletter
\renewcommand{\@noticestring}{}
\makeatother
\nolinenumbers
\usepackage[utf8]{inputenc}
\usepackage[T1]{fontenc}
\usepackage{hyperref}
\usepackage{url}
\usepackage{booktabs}
\usepackage{array}
\usepackage{multirow}
\usepackage{amsfonts}
\usepackage{nicefrac}
\usepackage{microtype}
\usepackage{xcolor}
\usepackage{graphicx}
\usepackage{amsmath}
\usepackage{float}
\usepackage{placeins}

\title{Features have life history. \\ And we should care.}

\author{%
  Philipp Stecher \\
  University of T\"ubingen \\
  \texttt{philippmstecher@gmail.com}
  \And
  Sandro Radovanovi\'c \\
  University of Belgrade
  \And
  Vlasta Sikimi\'c \\
  Eindhoven University of Technology
  \And
  Reinhard Kahle \\
  University of T\"ubingen
}

\begin{document}
\maketitle

\begin{abstract}
Features in language models have life history: they emerge, persist, and die during training, yet the importance of that history remains largely unexplored. We find evidence of a persistent representational backbone, which we identify in Pythia-160M and -410M as the carrier scaffold: ${\sim}50$ sparse features with stable life histories, around which the model's representational structure organises. It has four properties. \emph{(i)}~\emph{It assembles early:} features emerge, die, and reorganise ${\sim}40\!\times$ faster in the first $1\%$ of training than afterwards, and the scaffold is already largely fixed by then. \emph{(ii)}~\emph{It is load-bearing:} joint cross-layer ablation identifies the carriers as far more load-bearing than any count-matched non-scaffold population, a gap invisible to per-firing single-feature methods. \emph{(iii)}~\emph{Function precedes direction:} which features will become carriers is already predictable from training-onset firing patterns alone, correctly distinguishing future carriers from non-carriers in $4$ of $5$ cases, before the geometry has settled. \emph{(iv)}~\emph{It seeds subsequent development:} by the end of training, scaffold carriers have recruited $64\%$ of all active features into the scaffold hierarchy. Life history is consistent with a two-phase account of training: selection appears to largely determine the scaffold in the first $1\%$; the remaining $99\%$ appears to calibrate geometry around a substrate already set.
\end{abstract}

\section{Introduction}
\label{sec:intro}

Training is not uniform. The early phase of learning is qualitatively different from what follows: features are born and die at high rates, representations shift rapidly, and the model's internal structure reorganises faster than at any later point. By the time training ends, this turbulence has resolved into something stable, but the structure that stability settled around is not arbitrary. Some features form early, hold their position throughout, and end up carrying a disproportionate share of the model's representational load. Others arrive later, occupy similar positions at convergence, and carry far less. A snapshot cannot tell them apart, yet understanding how that difference arises is an interpretability question with direct consequences for model debugging, editing, and safety.

We propose feature life history as the unit of analysis and operationalise it with sparse autoencoders (SAEs) \citep{bricken2023monosemanticity, gao2024scaling}, which decompose the model's internal activations into a sparse set of interpretable features. A feature's life history tracks two properties across training: what it responds to (its firing identity) and how it influences predictions (its decoder direction). These calibrate on very different timescales: firing identity stabilises early, anchored to training data that is fixed from the start; decoder direction continues to be worked out across the bulk of training. Identifying each feature by firing identity rather than by its position in activation space gives us both a way to track features across checkpoints and a natural hierarchy in which broader features contain the firing patterns of narrower, more specialised ones. Tracking both properties across checkpoints lets us identify a small population of features whose firing identity has been stable since the opening phase of training and which have organised the majority of the model's representational structure around them by convergence. We call this population the carrier scaffold.

We train SAEs at seven points during training on two scales (Pythia-160M and Pythia-410M \citep{biderman2023pythia}) and follow each feature's birth, persistence, and position in this hierarchy. Recent work tracks SAE features across checkpoints \citep{yang2024saetrack, ge2025evolution, bayazit2025crosscoding, inaba2025howllms}; \citet{aoyama2025induction} predict induction-head emergence training-time, closest in spirit to our scaffold predictor. Prior approaches embed identity architecturally, via shared crosscoder encoders \citep{ge2025evolution, bayazit2025crosscoding} or warm-start weight inheritance \citep{yang2024saetrack}, or forgo individual tracking altogether \citep{inaba2025howllms}; we instead train independent SAEs at each checkpoint and match post-hoc by activation-profile correlation. Where prior work reports what features form and when, we additionally test whether the early-formed population is load-bearing, show that its membership is predictable from firing patterns before representations converge, and trace how it seeds the mature network's representational structure.

\textbf{Contributions.}
The central methodological contribution is \emph{feature life history} as a unit of analysis. A feature broad at convergence that has been broad since step~$1$k is a different scientific object from one that is broad at convergence due to noise in the SAE training: the former is part of the load-bearing scaffold; the latter is not. Tracking each feature's firing identity and decoder direction across checkpoints reveals a temporal split, invisible to any single-snapshot analysis, that makes the carrier scaffold recoverable.
\begin{enumerate}
  \item \textbf{The scaffold is born early (\S\ref{sec:r-substrate})}: the per-step feature birth rate collapses ${\sim}40\!\times$ in the first ${\sim}1\%$ of training, and most carriers are already identifiable at that point, before the model can reliably predict; a small persistent population forms one connected cross-layer structure, replicating at both scales.
  \item \textbf{The scaffold is load-bearing (\S\ref{sec:r-pivot})}: jointly ablating persistent carriers incurs far greater cross-entropy cost than removing any count-matched non-scaffold population at the same structural tier, a gap invisible to per-firing single-feature methods and emergent only under the joint cross-layer intervention; replicated at both scales.
  \item \textbf{Function precedes direction (\S\ref{sec:r-predictive})}: a logistic predictor trained on step-$1$k firing patterns correctly identifies future scaffold members in $4$ of $5$ pairwise comparisons before decoder directions have stabilised, with $9.3\!\times$ enrichment at the top operating point.
  \item \textbf{The scaffold seeds subsequent feature development (\S\ref{sec:r-seeding})}: scaffold carriers recruit subsequently born features via hierarchical connections, organising $64\%$ of active features at Pythia-160M ($53\%$ at Pythia-410M) into the scaffold hierarchy by convergence, invisible to trained-snapshot analysis and recoverable only by tracking the scaffold's growth across training.
\end{enumerate}

\section{Method}
\label{sec:method}

We track SAE features across training checkpoints of two model scales, identify a persistent carrier population, and test whether it is load-bearing. We instrument two Pythia scales across seven checkpoints, train one SAE per (layer, checkpoint) pair, match features across time to build life histories and a cross-layer DAG, and measure load via joint cross-layer ablation and a decoder transplantation control. A training-time predictor quantifies how early the scaffold is identifiable from function alone. Full methodological details are in App.~\ref{app:C}.

\subsection{Models, data, and checkpoints}

We instrument Pythia-160M and Pythia-410M \citep{biderman2023pythia} at seven training checkpoints: steps 0, 1k, 10k, 40k, 80k, 120k, and 143k (full schedule in Table~\ref{tab:setup}, App.~\ref{app:C}). Steps 0--10k cover ${\sim}7\%$ of training by wall-clock progress but concentrate the majority of life events; step~143k is the convergence checkpoint. We instrument four residual-stream layers per scale at approximately uniform fractional depth: layers 2, 5, 8, 11 at Pythia-160M (${\sim}18\%$, $45\%$, $73\%$, $100\%$) and layers 4, 11, 17, 23 at Pythia-410M (${\sim}17\%$, $46\%$, $71\%$, $96\%$). All SAE training, feature matching, and ablation experiments use a held-out slice of The Pile \citep{gao2020pile} not seen during Pythia pretraining; feature-level statistics are computed on a canonical validation sample $S^*$ of $20{,}000$ token positions drawn from this slice (seed $= 0$; App.~\ref{app:C}).

\subsection{Sparse autoencoders}

At every (checkpoint, layer) pair we train a top-$k$ SAE \citep{gao2024scaling, bricken2023monosemanticity, sharkey2022superposition} on residual-stream activations:
\[
  \hat{x} = W_{\mathrm{dec}}\,\mathrm{TopK}(W_{\mathrm{enc}}\,x + b_{\mathrm{enc}}) + b_{\mathrm{pre}},
  \qquad
  \mathcal{L} = \|x - \hat{x}\|_2^2,
\]
with $k{=}32$ simultaneously active features, expansion factor $64$, decoder columns unit-normalised at each step, and Adam ($\eta{=}3{\times}10^{-4}$, batch size $64$) for $4{,}000$ steps on the Pile validation slice. A feature is alive at a checkpoint if its activation exceeds $10^{-6}$ on ${\geq}0.5\%$ of $S^*$ samples. SAE reconstruction quality (variance explained, cosine similarity to the original residual stream) is reported in App.~\ref{app:C}. A robustness control with a higher-budget SAE (more samples, more steps) is in App.~\ref{app:C}.

\subsection{Feature tracking: matching, life events, and survival}

At each consecutive checkpoint pair $(t, t')$, features are matched by Hungarian assignment \citep{kuhn1955hungarian} that maximises total similarity:
\[
  M^* = \arg\max_{M \in \mathcal{M}} \sum_{(u,v) \in M} r(u, v),
  \qquad
  r(u,v) = \frac{\sum_{s \in S^*}(a_u^t(s)-\bar{a}_u^t)(a_v^{t'}(s)-\bar{a}_v^{t'})}
                {\|\mathbf{a}_u^t - \bar{a}_u^t\|\,\|\mathbf{a}_v^{t'} - \bar{a}_v^{t'}\|},
\]
where $a_u^t(s)$ is feature $u$'s activation on sample $s$ at checkpoint $t$ and $\mathcal{M}$ is the set of one-to-one matchings; a match is accepted only when $r \geq 0.5$. Each transition is classified as \emph{stable} (matched), \emph{born} (unmatched, newly alive), or \emph{died} (unmatched, no longer alive). A feature lineage is the maximal chain of stable matches from birth. Event types (Assembly, Decay, Task-general, Abstraction, Differentiation) follow the RDT taxonomy of \citet{stecher2025birthloss}, with CI replacing the label-derived selectivity indices of the original (App.~\ref{app:C}).

Survival to step~$143$k is computed as the fraction of features within a firing-entropy quartile at step~$1$k whose lineage persists to the final checkpoint; quartiles are defined over all alive features at step~$1$k within each scale. Directional precedence in Phase~I event-rate sequences (births, abstractions, differentiations at each layer) is assessed via lag-$1$ Pearson correlations on log-transformed event counts (App.~\ref{app:C}). Robustness controls (independent-initialisation replication; higher-budget SAE) are in App.~\ref{app:C}.

\subsection{Persistent carriers and the cross-layer DAG}

A lineage is a \emph{persistent carrier} if it is stable in $\geq 4$ of the $6$ post-initialisation snapshots (steps 1k--143k). The Containment Index of feature $u$ with respect to feature $v$ is
\[
  \mathrm{CI}(u,v) \;=\; \frac{|T_u \cap T_v|}{|T_u|},
\]
where $T_u$ is the set of $S^*$ tokens on which $u$ activates above threshold. CI identifies feature hierarchies from nested firing patterns alone, without semantic labels. The cross-layer DAG has persistent carriers as nodes; an edge from $u$ at layer $\ell$ to $v$ at layer $\ell' > \ell$ is drawn when the subset fraction $|T_u \cap T_v|/|T_v| \geq \tau$ (calibrated per snapshot to control null-edge density), the co-firing lift exceeds $2.0$, and the edge survives $\geq 80\%$ of bootstrap resamples; full construction details in App.~\ref{app:C}.

\emph{Sub-DAG coverage} at step $t$ is the fraction of all active features at $t$ that are reachable from at least one persistent carrier through any sequence of edges in the cross-layer DAG (BFS transitive closure from the carrier seed set). \emph{Firing breadth} of a feature is the fraction of $S^*$ tokens on which its activation exceeds the alive threshold.

\subsection{Ablation load and functional decomposition}

Per-firing single-feature impact \citep{makelov2024principled, marks2025sparse} (the $\Delta$CE when one feature's reconstruction is zeroed) is the field-default importance readout. We instead read load at the population level: for target population $P$, we simultaneously zero every feature in $P$ across all four instrumented layers, then measure aggregate $\Delta\mathrm{CE}$ (nats above SAE roundtrip baseline) on a held-out validation pass ($95\%$ bootstrap CI, $n{=}1{,}000$ resamples of $32$ evaluation chunks). Three populations form the comparison: (1)~persistent carriers ($51$ at 160M, $49$ at 410M); (2)~non-persistent roots (NCR): features at the same layer and CI tier but stable across fewer than four snapshots, count-matched to carriers; (3)~matched leaves: CI-tier-0 features breadth-matched by layer and firing-rate. NCR isolates the contribution of temporal persistence from structural position; leaves establish the near-zero baseline for non-hierarchical features.

To isolate decoder direction's contribution to load, we replace each persistent carrier's step-$143$k decoder column with its step-$1$k value while leaving all other model weights unchanged, then measure $\Delta\mathrm{CE}$ on the same validation pass. A complementary condition replaces only the encoder row. The difference between full ablation and transplantation costs quantifies how much load derives from decoder-direction calibration across Phases~II--III versus from firing-pattern identity fixed in Phase~I.

Feature function is its firing-pattern identity: the per-token activation distribution across $S^*$ \citep{shea2018representation, baker2022three, stecher2026scaffolded}. Feature direction is its decoder column. Step-to-step convergence is measured by Pearson $r$ of activation profiles (function) and cosine similarity of decoder columns (direction) to their step-$143$k values; full curves are in App.~\ref{app:D}. For \S\ref{sec:r-predictive}, a logistic regression with five-fold cross-validation is fit on six step-$1$k firing-space properties to predict step-$143$k persistent-carrier identity. Leave-one-layer-out cross-validation additionally assesses generalisation across layers. Full predictor covariates, coefficient estimates, and operating-point statistics are in App.~\ref{app:F}.
\section{Results}
\label{sec:results}
\vspace{-6pt}

Feature life history offers a complementary unit of analysis for mechanistic interpretability.
The four sections below provide evidence that the scaffold is born early
(\S\ref{sec:r-substrate}), load-bearing (\S\ref{sec:r-pivot}), predictable from
function alone (\S\ref{sec:r-predictive}), and organisationally dominant by
convergence (\S\ref{sec:r-seeding}).

\FloatBarrier
\subsection{The scaffold is born early}
\label{sec:r-substrate}

\begin{figure}[H]
  \centering
  \includegraphics[width=\linewidth]{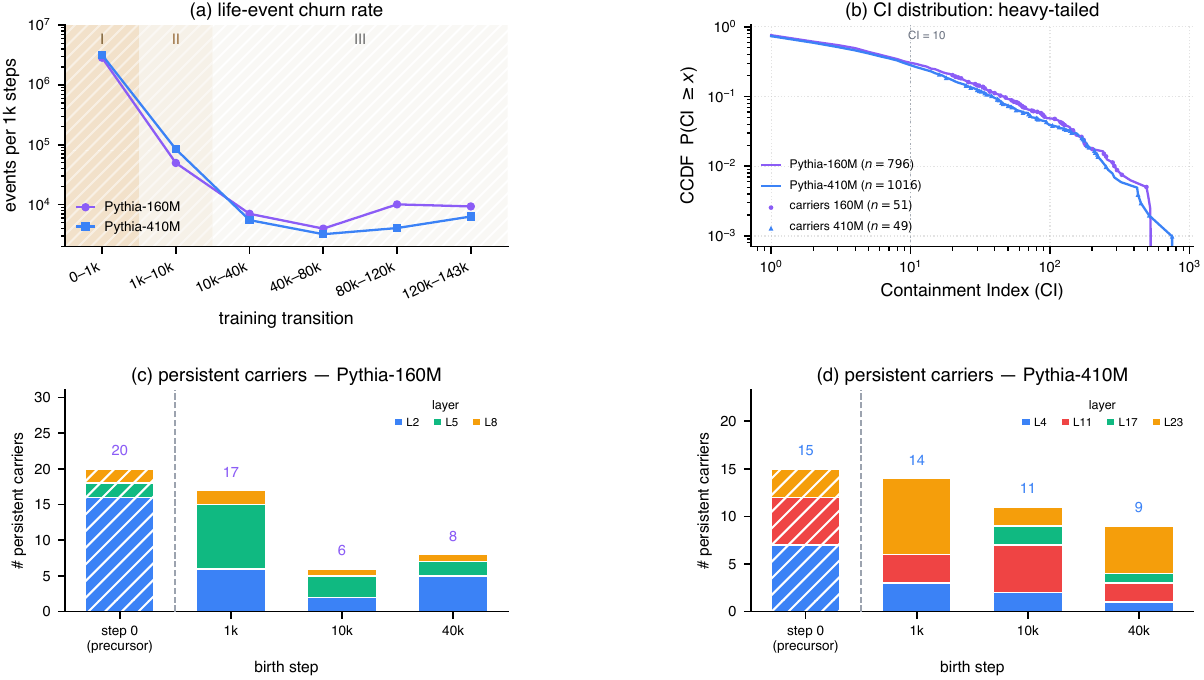}
  \caption{The scaffold's assembly dynamics and its organisational reach. \textbf{(a)} Per-step feature life-event rate at Pythia-160M and -410M across Phase~I Genesis (steps $0$--$1$k, dark shading), Phase~II Assembly ($1$k--$10$k, lighter), and Phase~III Refinement ($10$k--$143$k). Event types: As-E Assembly (birth), De-E Decay (death), Tg-E Task-general (broadening to root tier), Ab-E Abstraction (broadening to mid tier), Di-E Differentiation (narrowing to leaf tier); full definitions in App.~\ref{app:C}. \textbf{(b)} CCDF of the Containment Index (CI) on log-log axes at step~$143$k; approximately linear over CI $\in [5,80]$ ($\alpha\!\approx\!-0.80$). Features with CI $\gg 10$ are the persistent carrier seeds. \textbf{(c)} Persistent-carrier birth-step distribution at Pythia-160M (bars stacked by layer: L2 blue, L5 green, L8 amber); hatched bar: pre-training precursors already weakly active at step~$0$ (App.~\ref{app:C}). L11 contributes zero carriers (substrate failure; App.~\ref{app:E}). \textbf{(d)} Same at Pythia-410M (L4 blue, L11 red, L17 green, L23 amber). Roughly $40\%$ (160M) and $31\%$ (410M) of carriers are pre-training precursors.}
  \label{fig:1}
\end{figure}

We divide training into three phases: Phase~I Genesis (steps $0$--$1$k, the first ${\sim}1\%$ of training), Phase~II Assembly (steps $1$k--$10$k), and Phase~III Refinement (steps $10$k--$143$k). The per-step feature birth rate falls ${\sim}40\!\times$ across Phase~I and a further ${\sim}10\!\times$ over the remainder (Fig.~\ref{fig:1}a; App.~\ref{app:C}), compressing representational work into the opening phase. The Containment Index (CI) distribution at the end of training is heavy-tailed (Fig.~\ref{fig:1}b; approximately linear on log-log axes, slope $\alpha\!\approx\!-0.80$ over CI $\in [5, 80]$): a small number of features contain the firing patterns of many others, while the majority are narrow and uncontained.

Survival to step~$143$k is not random. Features in the top firing-entropy quartile at step~$1$k survive at $43\%$ at Pythia-160M vs.\ $0.6\%$ for bottom-quartile features (a $72\!\times$ survival advantage; Pythia-410M: $38\%$ vs.\ $0.4\%$; App.~\ref{app:C}). A lag-$1$ Granger test on Phase~I event sequences is consistent with directional precedence: birth events precede abstraction events ($F \in [8.4, 14.1]$, $p < 0.01$ at all four Pythia-160M layers; App.~\ref{app:C}).

The scaffold's semantic content and containment structure are detailed in App.~\ref{app:A} and~\ref{app:B}: categories emerge from co-firing statistics rather than imposed labels, and $99.6\%$ of scaffold edges link features by overlapping context rather than strict token-set inclusion. Scaffold carriers span a content gradient with depth: byte-boundary infrastructure and morphological features at early layers, discourse markers at mid layers, and high-level discourse abstractions at later layers.
The features in the right tail of this distribution are the persistent carriers: SAE features whose chain-matched lineage is stable in at least $4$ of the $6$ post-initialisation snapshots (App.~\ref{app:C}). $51$ at Pythia-160M and $49$ at Pythia-410M, they form one connected cross-layer object spanning all four layers at Pythia-410M and three of four at Pythia-160M (L11 absent due to substrate failure; App.~\ref{app:E}; Fig.~\ref{fig:1}c,d). Throughout, ``scaffold'' refers to this population, and the cross-layer DAG (nodes: persistent carriers; edges: token-set containment between layers, by the Containment Index) is the graph among them. By step~$1$k, $73\%$ of 160M and $59\%$ of 410M carriers ($37/51$ and $29/49$) are already present (Fig.~\ref{fig:1}c,d), while validation cross-entropy still exceeds $5$ nats at Pythia-160M (final: ${\sim}2.9$ nats): the model appears to commit to its scaffold long before it can predict competently. About $35\%$ (160M) and $31\%$ (410M) of eventual carriers are already weakly active at step~$0$, before any gradient update (App.~\ref{app:C}), so Phase~I partly amplifies latent structure from the random weight matrix rather than building the scaffold entirely from scratch.

Semantic labels and blind-LLM validation for all $51$ carriers are in App.~\ref{app:A}; tier lifespan, firing breadth, and semantic scope gradient across roots, mids, and leaves in App.~\ref{app:B}. These broad-born survivors are not isolated: their cross-layer containment edges already recruit subsequently born features from step~$1$k onward, with decoder directions on scaffold edges aligned an order of magnitude above a random-pair null (App.~\ref{app:D}).

\FloatBarrier
\subsection{The scaffold carries load}
\label{sec:r-pivot}

The cross-layer DAG was constructed from token-set containment alone, with no reference to ablation impact. We test whether the persistent carriers are load-bearing using joint cross-layer ablation (\S\ref{sec:method}): simultaneously replacing every feature in a target population with the batch mean across all four layers and measuring the aggregate $\Delta$CE on a held-out validation pass above the SAE roundtrip baseline.

\begin{figure}[H]
  \centering
  \includegraphics[width=\linewidth]{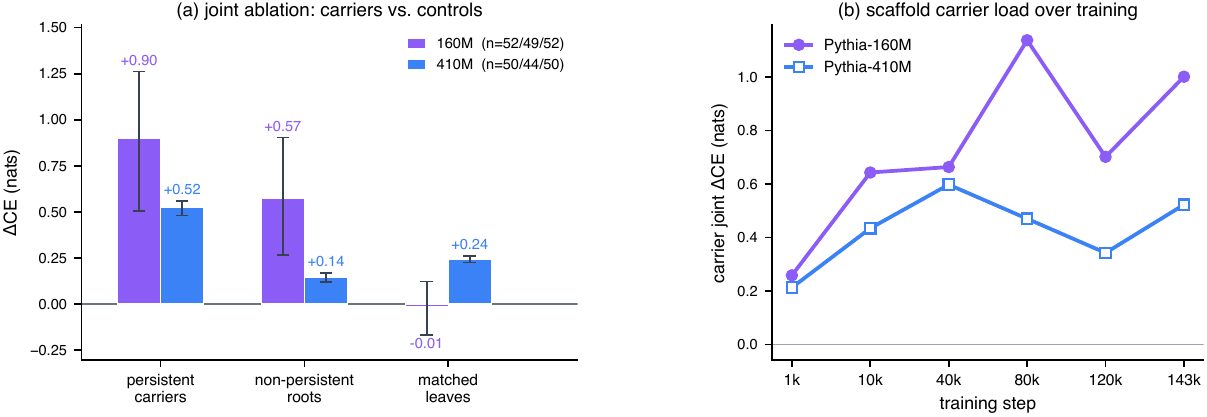}
  \caption{Joint cross-layer ablation quantifies the scaffold's load. \textbf{(a)} Joint $\Delta$CE above SAE roundtrip baseline for three count-matched groups, both scales shown side by side: persistent carriers (purple/blue), non-persistent roots (lavender/light blue), matched leaves (grey). Error bars show $95\%$ bootstrap CIs ($n{=}1{,}000$ resamples). Per-firing $\Delta$CE is tier-invariant (App.~\ref{app:D}). \textbf{(b)} Carrier joint $\Delta$CE over training for both scales. A transient dip at step~$120$k at Pythia-160M coincides with the L11 substrate reorganisation (App.~\ref{app:E}).}
  \label{fig:2}
\end{figure}

At Pythia-160M, jointly ablating all $51$ persistent carriers costs $+0.90$ nats above the SAE roundtrip baseline ($95\%$ bootstrap CI $[+0.51, +1.26]$, $n{=}1{,}000$ resamples of $32$ eval chunks). Ablating $49$ count-matched non-persistent roots (NCR; same layer and CI tier, chain-matched across fewer than four snapshots) costs $+0.57$ nats (CI $[+0.27, +0.90]$). Ablating $52$ breadth-matched leaves costs $-0.01$ nats (CI $[-0.17, +0.12]$); Fig.~\ref{fig:2}a.

The carrier-to-leaf gap is ${\sim}91\!\times$: $+0.90$ nats vs.\ $-0.01$ nats. This gap is invisible to the standard per-firing single-feature diagnostic, which assigns statistically indistinguishable importance across root, mid, and leaf tiers (App.~\ref{app:D}); it emerges only under the cross-layer joint intervention introduced here.

The carrier--NCR gap ($+0.33$ nats; NCR has three fewer features) isolates the contribution of temporal persistence over structural position: two features can occupy the same CI tier at the final checkpoint yet differ in ablation load by ${\sim}1.6\!\times$, on whether that position was held stably throughout training. For tier anatomy and per-firing tier-invariance see App.~\ref{app:B}.

The pattern replicates at Pythia-410M: carriers $+0.52$ nats (CI $[+0.48, +0.56]$), NCR $+0.14$ nats (CI $[+0.12, +0.17]$), leaves $+0.24$ nats (CI $[+0.23, +0.26]$; Fig.~\ref{fig:2}a). At both scales the carrier load is dominated by early layers; at Pythia-160M no carriers survive at L11 (App.~\ref{app:E}). Carrier load is not static: joint $\Delta$CE rises monotonically across training checkpoints at Pythia-160M and remains elevated throughout at Pythia-410M (Fig.~\ref{fig:2}b), consistent with the scaffold accumulating load as later training calibrates decoder directions around its fixed functions.

A temporal split underlies the load. Every persistent carrier has two distinguishable properties: its \emph{function} (per-token activation distribution) and its \emph{direction} (decoder column). Function converges substantially faster: correlation with the step-$143$k firing pattern is already high at step~$1$k, whereas decoder-column cosine begins near-random and calibrates across Phase~II--III. A transplantation control isolates direction's contribution: replacing each carrier's step-$143$k decoder column with its step-$1$k value while leaving all other weights unchanged costs $+0.96$ nats at Pythia-160M and $+0.61$ nats at Pythia-410M (nearly matching full ablation). Replacing only encoder rows costs $+0.58$ nats (160M) and $+0.44$ nats (410M), suggesting that decoder direction calibration across Phase~II--III accounts for the dominant share of the scaffold's load.

\begin{table}[h]
\centering
\caption{Summary of key metrics at both scales. $\Delta$CE values are joint cross-layer ablation costs in nats above SAE roundtrip baseline ($95\%$ bootstrap CI, $n{=}1{,}000$). Sub-DAG coverage is the fraction of active features inside the carrier sub-DAG at step~$143$k.}
\label{tab:results_summary}
\resizebox{\linewidth}{!}{%
\begin{tabular}{lcccccc}
\toprule
Model & Carriers & Born $\leq$ step~1k & $\Delta$CE carriers & $\Delta$CE NCR & AUC (step~1k) & Sub-DAG \\
\midrule
Pythia-160M & 51 & 73\% (37/51) & $+0.90$ [$+0.51$,$+1.26$] & $+0.57$ [$+0.27$,$+0.90$] & 0.80 & 64\% \\
Pythia-410M & 49 & 59\% (29/49) & $+0.52$ [$+0.48$,$+0.56$] & $+0.14$ [$+0.12$,$+0.17$] & 0.83 & 53\% \\
\bottomrule
\end{tabular}%
}
\end{table}

Load within the scaffold is not uniformly distributed: byte-boundary infrastructure carriers dominate the joint ablation cost over the semantically legible subset (App.~\ref{app:A}).

The temporal split suggests that scaffold formation may be predictable from function alone, before direction converges (\S\ref{sec:r-predictive}).

\FloatBarrier
\subsection{Function precedes direction}
\label{sec:r-predictive}

Firing identity is largely settled at step~$1$k while decoder direction has barely begun to calibrate, opening an audit window a simple predictor can exploit.

\begin{figure}[H]
  \centering
  \includegraphics[width=\linewidth]{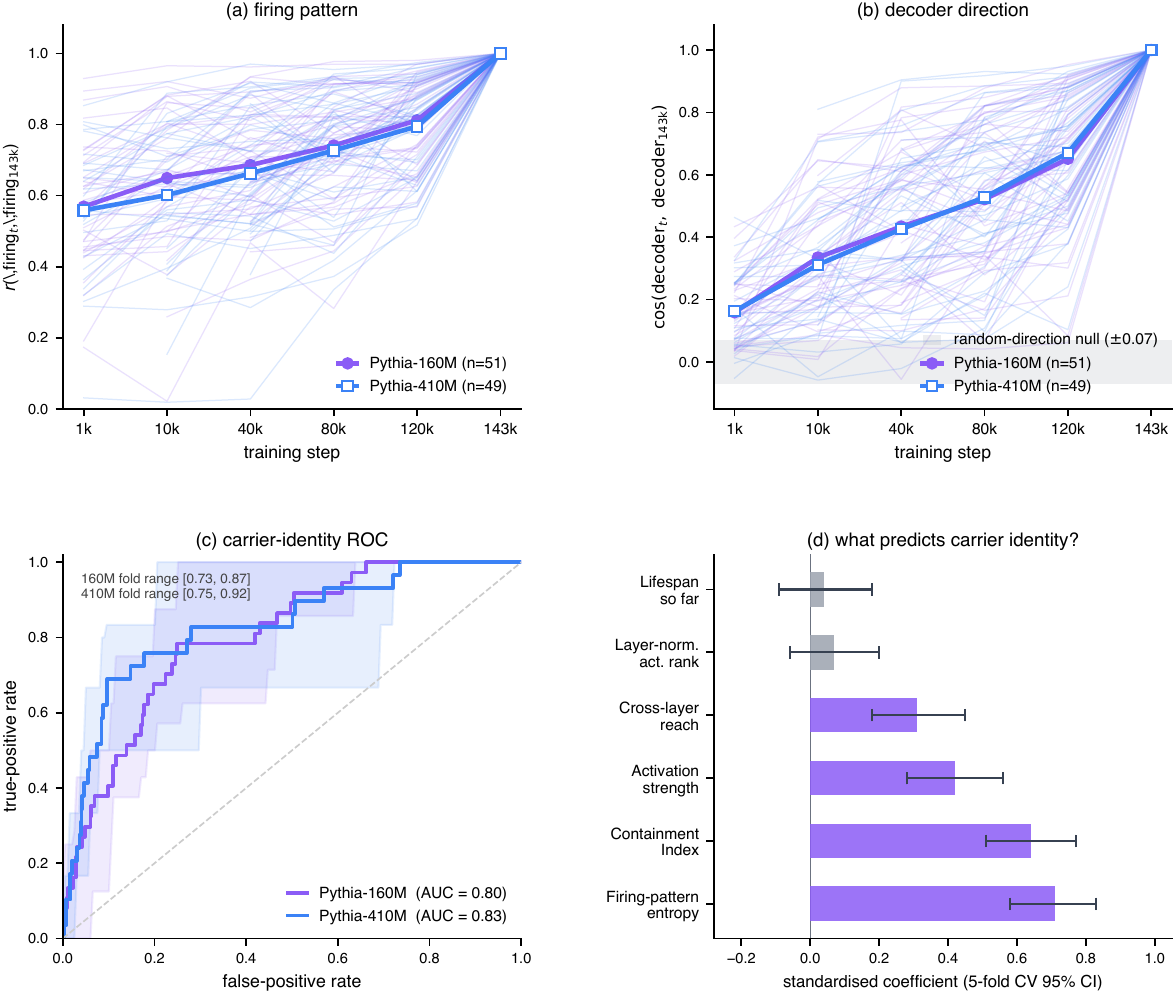}
  \caption{Function settles early; direction calibrates late; the gap makes the scaffold legible at step~$1$k. \textbf{(a)} Per-carrier Pearson $r$ to step-$143$k firing pattern ($n = 51/49$; faint lines: individual carriers; bold: per-scale means). Identity is largely fixed within Phase~I. \textbf{(b)} Per-carrier cosine to step-$143$k decoder column. Direction starts near-random at step~$1$k and rises across Phases~II--III. \textbf{(c)} Five-fold CV ROC predicting step-$143$k carrier identity from step-$1$k features: Pythia-160M (purple, AUC $= 0.80$, fold range $[0.73, 0.87]$) and Pythia-410M (blue, AUC $= 0.83$, $[0.75, 0.92]$); shaded bands span per-fold min/max TPR. \textbf{(d)} Standardised logistic-regression coefficients (5-fold CV 95\% CI). Firing breadth and CI at step~$1$k dominate; max activation, conditional mean, activation variance, and root flag are secondary.}
  \label{fig:3}
\end{figure}

Six step-$1$k firing-space properties predict step-$143$k persistent-carrier identity at five-fold cross-validated AUC $= 0.80$ at Pythia-160M and $0.83$ at Pythia-410M (Fig.~\ref{fig:3}c; standardised coefficients in Fig.~\ref{fig:3}d and App.~\ref{app:F}). At the top-$37$ operating point (matching $n_{\text{pos}}$), the predictor recovers $4$ of $37$ true carriers (precision $10.8\%$, a $9.3\!\times$ enrichment over the $1.2\%$ base rate; App.~\ref{app:F}). Firing-pattern entropy is the dominant predictor (standardised coefficient $+0.71$; Fig.~\ref{fig:3}d); an entropy-only single-feature predictor recovers the majority of this signal, with Containment Index, activation strength, and cross-layer reach contributing the remaining incremental lift to $0.80$ AUC. The classifier does not memorise layer identity: leave-one-layer-out AUC $\in [0.74, 0.79]$ across the four held-out layers (App.~\ref{app:F}).

The predictor succeeds because the diagnostic exploits what is already settled at step~$1$k: firing identity is largely fixed in the Phase I window; decoder direction is not. Scaffold membership therefore appears to be grounded in function rather than direction. The features identifiable from this early window are not merely predictable. They appear to be the organisational seeds around which the rest of the network differentiates, as the next section shows.

\FloatBarrier
\subsection{The scaffold seeds subsequent development}
\label{sec:r-seeding}

\begin{figure}[H]
  \centering
  \includegraphics[width=\linewidth]{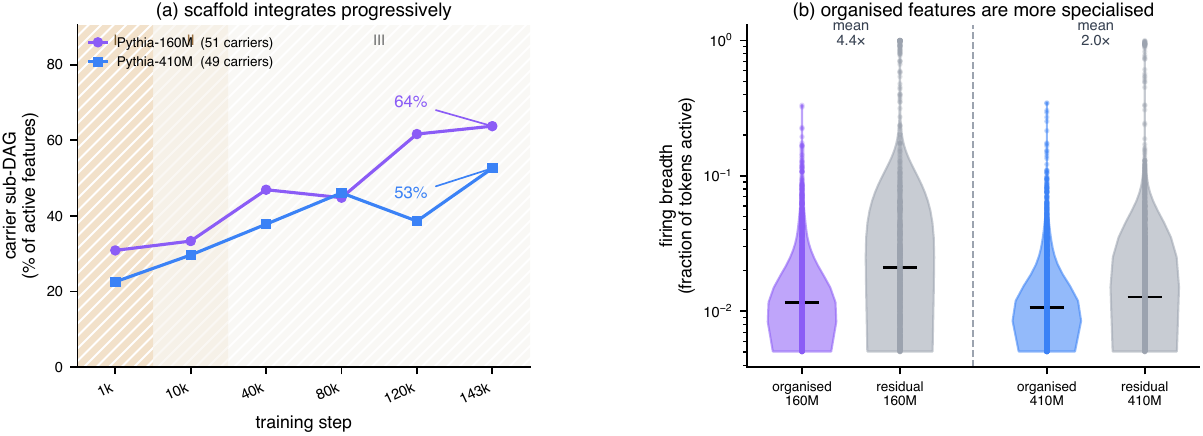}
  \caption{The scaffold organises two-thirds of the mature network through hierarchically scaffolded differentiation. \textbf{(a)} Carrier sub-DAG as percentage of active features at Pythia-160M (purple) and Pythia-410M (blue) across six training checkpoints on a log scale; phase shading as in Fig.~\ref{fig:1}c. \textbf{(b)} Firing breadth (fraction of tokens activating each feature) at step~$143$k for scaffold-organised features (inside carrier sub-DAG; purple/blue) versus unorganised residual features (grey), at both scales. Violins show full distributions; black bars mark medians. Mean breadth is $4.4\!\times$ lower inside than outside at Pythia-160M ($2.0\!\times$ at Pythia-410M). Grey dashed line separates the two scales.}
  \label{fig:4}
\end{figure}

A snapshot of the step-$143$k checkpoint sees $2{,}408$ active features at Pythia-160M: a flat list with no record of which features were born under which seed or when they were assimilated. The life-history framework converts that flat list into a structured partition: $51$ seeds locatable at step~$1$k, $1{,}484$ scaffold-organised descendants, and $873$ unorganised residual features that were never assimilated. The ablation and predictive results of \S\ref{sec:r-pivot}--\ref{sec:r-predictive} attach specifically to the seed population; without tracking, there is no way to distinguish seeds from descendants in the convergence snapshot, and the ${\sim}91\!\times$ joint-ablation load gap disappears into per-feature noise.

The persistent carriers are not a static load-bearing minority. Through Phase I and beyond, the scaffold grows: any newly born feature whose firing tokens are a subset of an existing carrier's is added as a descendant, deepening the hierarchy without displacing its seeds. This process of hierarchically scaffolded differentiation is consistent with the load concentration of \S\ref{sec:r-pivot}: the scaffold carries disproportionate joint-ablation impact and appears to function as the organisational spine around which the model's representational structure differentiates.

At step~$1$k, $37$ of the $51$ eventual carriers are present at Pythia-160M and their immediate containment descendants already account for $31\%$ of active features ($23\%$ at Pythia-410M). As training proceeds, new features born as more-specific sub-patterns of existing scaffold members are progressively organised under the scaffold hierarchy. By step~$143$k, the carrier sub-DAG encompasses $1{,}535$ of $2{,}408$ active features ($64\%$) at Pythia-160M and $2{,}164$ of $4{,}114$ ($53\%$) at Pythia-410M (Fig.~\ref{fig:4}a; per-layer tier trajectory in App.~\ref{app:B}, Table~\ref{tab:ci_trajectory}), a ${\sim}33$ percentage-point expansion consistent across scales.

The scaffold's organizational reach leaves a geometric signature. Features inside the sub-DAG fire with narrower token-set specificity (mean breadth $4.4\!\times$ lower than outside features at Pythia-160M; $2.0\!\times$ at Pythia-410M; Fig.~\ref{fig:4}b): recruitment by containment is associated with progressively more specific sub-patterns of the seed representations. The $873$ features outside the scaffold at Pythia-160M ($1{,}950$ at Pythia-410M) retain broad, undifferentiated firing patterns and fall outside the carrier hierarchy.

\clearpage
\section{Discussion}
\label{sec:discussion}

The results suggest that features have life history. Whether we should care depends on what changes when that history is known; four things do.

\emph{A selection mechanism becomes legible.} The function-precedes-direction asymmetry is consistent with a two-phase optimisation regime. In the early window, the training distribution may act as a selector: broadly-firing features occupy more token contexts, accumulate gradient signal across positions, and are less likely to go dark when the corpus distribution shifts. Data and task pressure appear to select which representational niches survive. Over the remaining training, we propose that gradient descent calibrates each surviving niche's decoder column toward the directions that best express its fixed function in the output space. Firing identity appears to be the outcome of selection; decoder direction appears to be the outcome of calibration around a fixed substrate. The temporal ordering is therefore expected rather than surprising, and the load asymmetry follows directly: scaffold membership (the outcome of selection), not any individual feature's direction, carries the load. This may explain why the asymmetry is invisible to per-firing single-feature methods and emerges only under joint intervention.

\emph{The unit interpretability has been operating on acquires a training-time grounding.} Sparse feature circuits \citep{marks2025sparse} and per-feature attribution methods \citep{meng2022rome, makelov2024principled} decompose the scaffold's joint load without observing it: the carrier--NCR load gap is invisible to them because per-firing impact is tier-uniform, and the structural object joint cross-layer ablation identifies is built on firing histories no trained-snapshot method records. Three practical consequences follow. Circuits assembled at the trained snapshot may benefit from being cross-checked against the scaffold population, since features outside it sit on calibrated directions but a non-load-bearing substrate. We predict fine-tuning rotates decoder directions while preserving scaffold function, testable per fine-tuning corpus. Training-time triage of feature populations appears viable from step-$1$k firing-space alone (\S\ref{sec:r-predictive}), long before the snapshot exists.

\emph{Three accounts of early-formed structure gain a measurable candidate substrate.} Critical learning periods \citep{achille2019critical, kleinman2024critical} predict irreversible damage from perturbation in the early window; the 160M-L11 substrate reorganisation (App.~\ref{app:E}) is consistent with a critical-period event at the SAE level, and the scaffold is the candidate substrate such disruption acts on. The lottery ticket hypothesis \citep{frankle2019lottery} names persistent subnetworks retrospectively; the persistent-carrier population is the training-time object those tickets may overlap with, auditable before any post-hoc pruning at AUC $= 0.80$. A third account is the broadly-firing $\to$ categorical $\to$ class-specific developmental ordering reported across CNN, ViT, and CCT architectures \citep{stecher2026scaffolded}; the persistent carrier scaffold is a candidate substrate for that ordering across domains, and the present work adds a load test the vision account does not perform.

Mechanistic interpretability has developed powerful tools for the trained snapshot. The life-history view complements them by surfacing a population whose load is established before convergence, traceable through firing-space, and invisible to any single-checkpoint analysis. The scaffold appears largely fixed by selection in the first $1\%$ of training; the remaining $99\%$ appears to calibrate geometry around a substrate that is already largely set. Interpretability that begins at convergence may be reading the product of that process more than its causes. Three directions follow directly: whether fine-tuning preserves scaffold function while rotating decoder directions; whether published circuits map onto the scaffold population; and whether scaffold size and load concentration generalise for large-scale language models.

\section{Limitations}
\label{sec:limitations}

We study Pythia at two scales on one training corpus with four layers per scale; whether scaffold size, load magnitude, and three-phase timing generalise to other architectures, training corpora, or model scales remains open. The evidence is correlational: ablation and prediction suggest that the scaffold is load-bearing and early-identifiable, but do not establish a causal mechanism for its formation. SAE coverage spans four of twelve (160M) and four of twenty-four (410M) residual-stream layers; denser layer sampling could reveal additional scaffold structure or alter the cross-layer connectivity reported here.

\bibliographystyle{unsrtnat}
\bibliography{references}

\newpage
\appendix

\section{Semantic Anatomy of the Carrier Scaffold}
\label{app:A}

Of the $51$ persistent carriers at Pythia-160M, $39$ are labelable from top-$8$ activating tokens; $82\%$ of those receive high-or-medium-confidence auto-interp labels, validated by blind LLM annotation ($90\%$ concept agreement; blind-validation protocol below). The load-bearing subset is dominated by byte-boundary infrastructure features rather than the semantically-interpretable discourse and morphological carriers, a result that rules out simple tokeniser-infrastructure confounds and localises the $2\!\times$ load concentration of \S\ref{sec:r-pivot} to a specific carrier subpopulation.

\paragraph{Auto-interp labels and confidence.}
For each of the $51$ carriers we extracted its top-$8$ activating tokens and logit-boost target class, and submitted both to an LLM for a short English label with high / medium / low / artefact confidence \citep{bills2023language, paulo2024automatically}. Table~\ref{tab:autointerp} summarises the distribution. $39$ of the $51$ carriers are labelable; $82\%$ of labelable carriers receive high-or-medium confidence labels. The layer profile tracks depth: L2 is heterogeneous (morphological endings, file-extension orthographics, discourse connectives, and a non-trivial tail of U+FFFD / CJK fragment noise); L5 consolidates into clearer discourse markers and semantic-topic clusters (licensing boilerplate, magnitude, causation); L8's six-carrier population is dominated by high-level discourse abstractions (paragraph-initial framing, modal predicates, temporal-narrative); L11 collapses to zero carriers (substrate failure; App.~\ref{app:E}).

\begin{table}[H]
  \centering
  \caption{Auto-interpretation confidence for the $51$ persistent carriers at Pythia-160M (step $143$k, $4\times$ expansion SAE). The artefact class consists of Unicode replacement characters, CJK fragment tokens, and HTML boundary separators concentrated at L2; they satisfy all scaffold criteria (persistent, root-level, monotonically sharpening decoder) but their semantics are byte-level infrastructure rather than discourse content.}
  \label{tab:autointerp}
  \small
  \begin{tabular}{lrrrr p{4.4cm}}
    \toprule
    Layer & Carriers & High & Medium & Low & Dominant categories \\
    \midrule
    L2  & 29 & 10 & 8 & 11 & Morphological, discourse connectives, CJK/U+FFFD noise \\
    L5  & 16 &  6 & 4 &  6 & Legal boilerplate, causal nouns, spatial particles \\
    L8  &  6 &  4 & 0 &  2 & Sentence-initial framing, modal predicates \\
    L11 &  0 &  0 & 0 &  0 & (substrate failure; App.~\ref{app:E}) \\
    \midrule
    Total & 51 & 20 & 12 & 19 & 82\% high/medium of 39 labelable \\
    \bottomrule
  \end{tabular}
\end{table}

\paragraph{Blind LLM validation ($90\%$ concept agreement).}
Twenty persistent carriers were drawn uniformly at random (seed 42) from the full $51$-carrier set across L2/L5/L8. Each carrier was presented to a second LLM with only its top-$8$ activating tokens and corresponding logit-boost values; the auto-interp label, confidence, rationale, and category were hidden. The second LLM provided a short English label and a high/medium/low confidence rating.
Blind-LLM vs.\ auto-interp concept agreement is $18/20$ ($90\%$). Confidence-level agreement (exact high/medium/low match) is $10/20$ ($50\%$). The direction of confidence disagreements is balanced: in $6$ of the $10$ cases the blind LLM is more confident than auto-interp, in $4$ auto-interp is more confident. The six blind-LLM-more-confident cases are dominated by U+FFFD- or CJK-dominated carriers that the blind LLM reads as coherent byte-boundary or multilingual-fragment detectors while auto-interp labels them as low-confidence noise. For example, L2 fid 13486 has all eight top tokens as U+FFFD replacement characters; the blind LLM labels this ``Unicode replacement / encoding-error characters'' at high confidence while auto-interp labels it ``Unicode replacement-character noise'' at low confidence. The two concept-level disagreements are carriers with genuinely mixed token sets. Under a stricter rubric that reclassifies mixed-token carriers as artefact-like, the high-or-medium rate of $82\%$ shifts to approximately $64\%$.

\paragraph{The semantic-confidence / load paradox.}
Within the $51$ carriers, ablation load is not uniformly distributed, and the partition is counterintuitive. Splitting by auto-interp confidence yields a $19$-carrier low/artefact subset (Unicode replacement characters, CJK fragments, HTML-adjacent separators; concentrated at L2 with 11, plus 6 at L5 and 2 at L8) and a $32$-carrier high-or-medium-confidence subset (discourse and morphological features across L2/L5/L8). Joint ablation of the low-confidence subset costs $\Delta\mathrm{CE} = +0.91$ nats ($95\%$ CI $[+0.06, +2.55]$); joint ablation of the high/medium-confidence subset costs $\Delta\mathrm{CE} = +0.001$ nats ($95\%$ CI $[-0.31, +0.17]$), indistinguishable from zero. The proportional shares would be $+0.37$ and $+0.63$ nats; the low-confidence subset carries $2.44\times$ its proportional share of the load. The blind LLM validation agrees with the auto-interp-based split on $9$ of $10$ items.
The semantically-interpretable discourse and morphological carriers are scaffold-shaped but carry near-zero load in isolation. The near-uninterpretable tokeniser-boundary infrastructure carriers carry the load. The paradox proactively answers the ``is this just tokenisation?'' objection: the scaffold result holds even if you exclude the byte-level infrastructure subset from the load argument; the $+1.00$-nat figure for all $51$ carriers is dominated by the infrastructure subset, but the whole $51$-carrier set satisfies every scaffold criterion.

\paragraph{Absorption vs.\ predictive nesting.}
\citet{chanin2024absorption} argue that sparsity-driven SAEs suffer feature absorption when underlying features form a taxonomic hierarchy. We compute for every edge $A \to B$ a subset ratio $|\mathrm{top\text{-}8}(A) \cap \mathrm{top\text{-}8}(B)| / 8$; an edge with subset ratio $\geq 0.5$ is counted as taxonomic, otherwise predictive. Of the $916$ containment edges in the $51$-carrier scaffold sub-DAG, $99.6\%$ are predictive (mean subset ratio $0.056$), vs.\ $95.2\%$ for the $3{,}409$ non-scaffold edges (odds ratio $\approx 11$, $p \ll 0.001$, Fisher exact). The scaffold is compositional rather than taxonomic: one feature predicts another's firing context rather than strictly subsuming it. A common-token filter (removing the $50$ most frequent Pile-val tokens) shifts the predictive fraction by less than $0.3$ pp.

\paragraph{Worked hierarchy trace: L8 fid 4795.}
Table~\ref{tab:worked_hierarchy} traces the highest-CI high-confidence carrier at step $143$k: L8 fid 4795 (``sentence-initial subordinating / framing connectives'': \emph{Although, Despite, If, Following, However}; CI $= 243$). The rows are $15$ direct descendants plus a $2$-hop expansion through nested root L8 fid 17144. The category column labels each level: paragraph-initial discourse framing (root), broad text-structural contexts (nested roots with diverse structural/boundary patterns), transitional mixed-context fragments (mids, no clean axis), and narrow local lexical content (leaves). The content gradient is broad-to-narrow but not strictly taxonomic: only three nested roots are genuine specialisations of the root's discourse-framing axis; the rest are contained for distributional reasons. Cross-layer reach: direct descendants in L2 ($4$), L5 ($4$), L8 ($7$).

\begin{table}[H]
\centering
\caption{Two-level trace of L8 fid 4795 at step $143$k. Rows sorted root $\to$ nested roots $\to$ mids $\to$ leaves (CI descending within level). Rows marked $\dagger$ are 2-hop descendants through nested root L8 fid 17144. The content gradient broad$\to$narrow is predictive not taxonomic ($99.6\%$ non-absorbed edges; see ``Absorption'' paragraph above).}
\label{tab:worked_hierarchy}
\small
\setlength{\tabcolsep}{3pt}
\renewcommand{\arraystretch}{1.15}
\begin{tabular}{@{}p{2.2cm} l l r p{3.4cm} p{3.4cm}@{}}
\toprule
category & level & feature & CI & top tokens & interpretation \\
\midrule
\bfseries paragraph-initial discourse framing & \textbf{root}  & (L8, 4795)  & 243 & \emph{If, however, Statistical, Following}    & sentence-initial framing markers \\
\midrule
\multirow[t]{7}{2.2cm}{\bfseries broad text-structural contexts}
  & nested root & (L2, 46887) & 119 & \emph{we, content, 6, \textbackslash n}       & paragraph-internal continuations          \\
  & nested root & (L2, 3039)  &  89 & \emph{/, N, younger, fine, videos}            & slash-delimited modifier fragments        \\
  & nested root & (L8, 17144) &  81 & \emph{Its, However, grid, a, del}             & sentence-initial concessive pronouns      \\
  & nested root & (L8, 39082) &  34 & \emph{other, Code, {,}, \textbackslash n}     & clause-connecting adjuncts                \\
  & nested root & (L5, 32927) &  21 & \emph{-, Society, the, \textbackslash n}      & dashed-list nominal references            \\
  & nested root & (L2, 48496) &  17 & \emph{\#, us, of, i, be}                      & hashtag \& function-word contexts         \\
  & nested root & (L2, 1062)  &  14 & \emph{\textbackslash n, on, \ldots, At, Fair} & paragraph-boundary markers                \\
\midrule
\multirow[t]{3}{2.2cm}{\bfseries transitional mixed-context fragments}
  & mid$^\dagger$ & (L2, 41638) &   9 & \emph{THE, -, ignoring, Partners}             & sentence-initial noisy capitals           \\
  & mid$^\dagger$ & (L5, 2609)  &   2 & \emph{GA, maintain, lead, ., really}          & action-verb context                       \\
  & mid           & (L5, 37931) &   1 & \emph{mosp, of, the, land, voted}             & sparse non-word $+$ function mix          \\
\midrule
\multirow[t]{9}{2.2cm}{\bfseries narrow local lexical content}
  & leaf           & (L5, 37889) &   0 & \emph{\textbackslash n, be, image, main}      & paragraph-start function tokens           \\
  & leaf$^\dagger$ & (L5, 9789)  &   0 & \emph{)., the, judge, \_, give}               & closing-punctuation context               \\
  & leaf           & (L5, 10272) &   0 & \emph{\ldots, crime, brew, played, crowned}   & past-tense narrative verbs                \\
  & leaf           & (L5, 12043) &   0 & \emph{array, Then, ', Left, she}              & sentence-start pronoun $+$ \emph{Then}    \\
  & leaf           & (L5, 17706) &   0 & \emph{now, TS, turnover, ress, clone}         & temporal adverb $+$ clipped nouns         \\
  & leaf           & (L8, 10322) &   0 & \emph{Partner, leader, \textbackslash n, are} & agent / leader nouns                      \\
  & leaf           & (L8, 15916) &   0 & \emph{thing, now, approximate, develop}       & near-future action verbs                  \\
  & leaf           & (L8, 41172) &   0 & \emph{You, and, of, ?), the}                  & second-person interrogative context       \\
  & leaf$^\dagger$ & (L8, 2169)  &   0 & \emph{Account, under, only, {,}, )}           & function-word closing context             \\
\bottomrule
\end{tabular}
\end{table}

\section{Tier Anatomy: What Roots, Mids, and Leaves Entail}
\label{app:B}

CI tier (root / mid / leaf by descendant count in the cross-layer DAG) strongly predicts a feature's life history and firing scope, but not its per-firing ablation impact. Roots and mids form one durability class; leaves are ephemeral. This dissociation is the quantitative backbone of the load-pivot claim in \S\ref{sec:r-pivot}: tier partitions features by representational role, but load only appears under joint cross-layer intervention (App.~\ref{app:D}).

\begin{figure}[H]
  \centering
  \includegraphics[width=\linewidth]{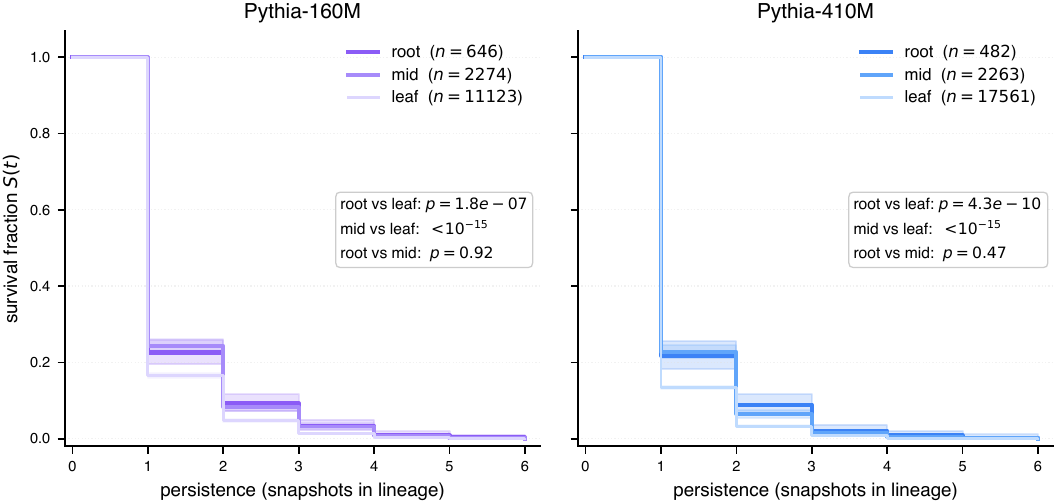}
  \caption{Kaplan--Meier survival by CI tier at Pythia-160M (left) and Pythia-410M (right), with Greenwood $95\%$ confidence bands. Roots and mids are statistically indistinguishable in durability ($p > 0.4$ at both scales); both massively outlive leaves ($p < 10^{-7}$). The tier partition that is invisible to per-firing ablation is a strong predictor of life history.}
  \label{fig:tier_lifespan}
\end{figure}

\paragraph{Lifespan by tier.}
Figure~\ref{fig:tier_lifespan} shows Kaplan--Meier survival curves for features stratified by birth tier at Pythia-160M and Pythia-410M. At 160M: root vs.\ leaf $p = 1.8 \times 10^{-7}$; mid vs.\ leaf $p < 10^{-15}$; root vs.\ mid $p = 0.92$. At 410M: root vs.\ leaf $p = 4.3 \times 10^{-10}$; mid vs.\ leaf $p < 10^{-15}$; root vs.\ mid $p = 0.47$. Roots and mids form one durability class; leaves are ephemeral. A feature born into a root or mid position will almost certainly persist to step $143$k; a feature born into a leaf position almost certainly will not.

\begin{figure}[H]
  \centering
  \includegraphics[width=0.55\linewidth]{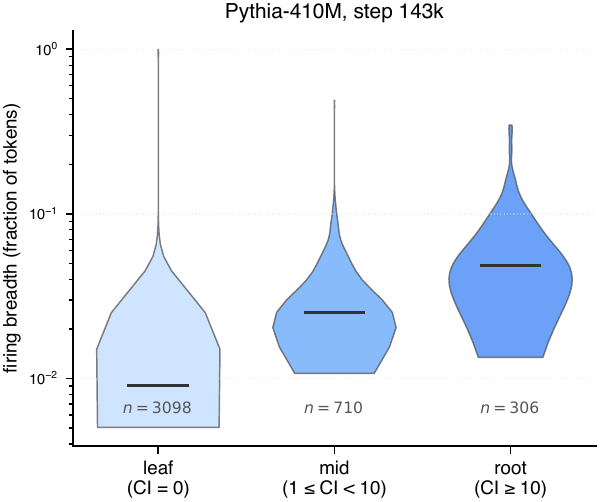}
  \caption{Firing breadth (fraction of validation tokens on which a feature activates) by CI tier at step $143$k, Pythia-410M; all four layers pooled ($n_\text{leaf}=3{,}098$, $n_\text{mid}=710$, $n_\text{root}=306$). Roots fire ${\sim}5\!\times$ more broadly than leaves at median (root $0.048$, leaf $0.009$); mids are intermediate. The same breadth axis is captured by the firing breadth predictor of \S\ref{sec:r-predictive}.}
  \label{fig:tier_breadth}
\end{figure}

\paragraph{Firing breadth by tier.}
Figure~\ref{fig:tier_breadth} shows firing breadth (fraction of validation tokens activating each feature) at step $143$k by CI tier, pooled across all four Pythia-410M layers. Roots fire roughly $5\!\times$ more broadly than leaves at median (root median $0.048$, leaf median $0.009$); mids are intermediate. The breadth axis is what the firing breadth predictor of \S\ref{sec:r-predictive} exploits.

\paragraph{Per-firing $\Delta$CE is tier-invariant.}
Despite the lifespan and breadth differences above, per-firing single-feature ablation impact is statistically indistinguishable across tiers (formal KS tests in App.~\ref{app:D}): load only emerges under joint cross-layer intervention.

\paragraph{CI trajectory per layer across training.}
Table~\ref{tab:ci_trajectory} gives the full per-snapshot trajectory for active features, roots, and scaffold sub-DAG coverage at each Pythia-160M layer. Root counts peak at step $1$k (L2: $221$, L5: $167$, L8: $176$, L11: $187$) and prune by roughly half to step $10$k, while sub-DAG coverage rises from $30.9\%$ at step $1$k to $63.7\%$ at step $143$k with a minor plateau around step $80$k. L11's $80\mathrm{k} \to 120\mathrm{k}$ collapse (active $612 \to 39$, roots $64 \to 0$) is the only qualitative departure; see App.~\ref{app:E} for the mechanistic account.

\begin{table}[H]
\centering
\caption{Per-layer active-feature and root counts across training at Pythia-160M, with cross-layer scaffold sub-DAG trajectory. Sub-DAG coverage grows overall from $31\%$ at step $1$k to $64\%$ at step $143$k. Root counts peak at step $1$k then prune by roughly half. Step $0$ is the random-init baseline.}
\label{tab:ci_trajectory}
\small
\setlength{\tabcolsep}{4pt}
\begin{tabular}{lrrrrrrr}
\toprule
layer & step 0 & step 1k & step 10k & step 40k & step 80k & step 120k & step 143k \\
\midrule
\multicolumn{8}{l}{\emph{active features}} \\
L2  & 1074 & 821 & 851 & 762 & 820 & 916 & 908 \\
L5  &  858 & 789 & 963 & 923 & 952 & 772 & 790 \\
L8  &  725 & 830 & 954 & 791 & 731 & 568 & 673 \\
L11 &  670 & 758 & 838 & 578 & 612 &  39 &  37 \\
\midrule
\multicolumn{8}{l}{\emph{roots (CI $\geq 10$)}} \\
L2  & 72 & 221 & 108 & 76 &  97 &  86 & 100 \\
L5  & 47 & 167 &  88 & 79 &  77 &  77 &  81 \\
L8  & 42 & 176 &  95 & 81 &  88 &  82 &  73 \\
L11 & 29 & 187 &  78 & 66 &  64 &   3 &   0 \\
\midrule
\multicolumn{8}{l}{\emph{cross-layer scaffold sub-DAG}} \\
total active                & 3327 & 3198 & 3606 & 3054 & 3115 & 2295 & 2408 \\
inside sub-DAG              &   78 &  987 & 1202 & 1433 & 1396 & 1415 & 1535 \\
coverage (\%)               &  2.3 & 30.9 & 33.3 & 46.9 & 44.8 & 61.7 & \textbf{63.7} \\
\bottomrule
\end{tabular}
\end{table}

\section{Methodological Details and Controls}
\label{app:C}

Full definitions, construction details, and robustness controls are collected here. Sections cover the persistent-carrier criterion, cross-layer DAG construction, the RDT event taxonomy with per-window counts, the lag-$1$ Granger test on Phase I sequences, broad-born survival curves, pre-existing structure at random initialisation, and seed and expansion-factor replication.

\begin{table}[H]
  \centering
  \caption{Experimental setup. Seven training checkpoints per scale (six post-initialisation + step~$0$); four residual-stream layers per scale at uniform depth spacing.}
  \label{tab:setup}
  \begin{tabular}{lll}
    \toprule
    Scale & Checkpoints (step) & Layers \\
    \midrule
    Pythia-160M & $0$, $1$k, $10$k, $40$k, $80$k, $120$k, $143$k & L2, L5, L8, L11 \\
    Pythia-410M & $0$, $1$k, $10$k, $40$k, $80$k, $120$k, $143$k & L4, L11, L17, L23 \\
    \bottomrule
  \end{tabular}
\end{table}

\paragraph{Persistent-carrier definition (full).} Following the criterion in \S\ref{sec:method}: a persistent carrier's lineage must be stable in at least $4$ of the $6$ post-initialisation snapshots at steps $1$k, $10$k, $40$k, $80$k, $120$k, $143$k (alive threshold $10^{-6}$ on ${\geq}0.5\%$ of validation samples; match threshold $r \geq 0.5$). At Pythia-160M this yields $51$ carriers across L2/L5/L8/L11; at Pythia-410M, $49$ across L4/L11/L17/L23.

\paragraph{Persistence-threshold sensitivity.} The canonical threshold ($\geq 4$ of $6$ post-initialisation snapshots) is the most conservative point at which the scaffold is definable. Tightening the threshold monotonically strengthens the early-commitment signal: at $\geq 5$ snapshots the 160M carrier count falls from $51$ to $43$ while the fraction born at step~$1$k rises from $73\%$ to $86\%$; at $\geq 6$ snapshots (present at every checkpoint) the count falls to $37$, all born at step~$1$k ($100\%$). The core finding is therefore not an artefact of threshold choice: stricter thresholds yield a smaller but more coherent scaffold, not a weaker one.

\begin{table}[H]
  \centering
  \caption{Persistence-threshold sensitivity at Pythia-160M. Stricter thresholds select a smaller but more cohesive carrier set with stronger early-commitment signal.}
  \label{tab:persistence_sensitivity}
  \small
  \begin{tabular}{lrrr}
    \toprule
    Threshold & $n$ carriers & \% born $\leq$ step~$1$k & \% step-$0$ precursors \\
    \midrule
    $\geq 4$ snapshots (canonical) & 51 & 73\% & 39\% \\
    $\geq 5$ snapshots             & 43 & 86\% & 47\% \\
    $\geq 6$ snapshots (maximum)   & 37 & 100\% & 54\% \\
    \bottomrule
  \end{tabular}
\end{table}

\paragraph{Cross-layer DAG construction.} For each ordered layer pair $(\ell, \ell')$ with $\ell < \ell'$ and each pair of carriers $(u, v)$ with $u$ at layer $\ell$ and $v$ at layer $\ell'$, we compute the subset fraction $s(u,v) = |T_u \cap T_v|/|T_v|$ and the co-firing lift $l(u,v) = s(u,v) / (b_u \cdot b_v)$ on a fixed validation slice, where $T_u, T_v$ are the firing-token sets and $b_u, b_v$ are their respective firing breadths. A directed edge $u \to v$ is drawn when (i)~$s(u,v) \geq \tau$, where $\tau$ is calibrated per snapshot to keep the expected null-edge count at or below $10$; (ii)~$l(u,v) > 2.0$; (iii)~$b_u > b_v$ (broader feature points to narrower); and (iv)~the edge survives in $\geq 80\%$ of $30$ bootstrap resamples of $S^*$. Transitive reduction is applied to yield the final DAG.

\paragraph{RDT mapping with CI.}\label{app:rdt}
We reproduce the RDT process definitions of \citet{stecher2025birthloss} with CI replacing the label-derived selectivity indices of the original. Thresholds: root $\geq 10$, mid $\in [1, 10)$, leaf $= 0$.
\begin{itemize}
  \item \textbf{Assembly} (As-E): feature unmatched at $T$, active at $T+1$ (``born'').
  \item \textbf{Decay} (De-E): feature active at $T$, unmatched at $T+1$ (``died'').
  \item \textbf{Task-general} (Tg-E): matched, and either born-at-root ($\mathrm{CI}_{T+1} \geq 10$, backward-$r < 0.2$) or matched-gaining ($z(\Delta\mathrm{CI}) \geq +0.5$, endpoint root).
  \item \textbf{Abstraction} (Ab-E): matched, born-at-mid or matched-gaining, endpoint mid.
  \item \textbf{Differentiation} (Di-E): matched, born-at-leaf or matched-losing, endpoint leaf.
\end{itemize}
Matched feature-events that do not cross any threshold do not receive a process label; they are ``matched background'' and are excluded from the process-composition analyses. The per-transition null $\mu, \sigma$ is estimated by permuting activation columns over position in the aligned sample $S^*$ per transition ($1{,}000$ permutations).

\paragraph{Per-token event counting.} Event counts in Table~\ref{tab:A1} are reported at the token level rather than the feature level. For each matched feature pair across a checkpoint transition, we record the set of validation tokens on which the event occurs: a Birth (As-E) token is one on which a confirmed-born feature activates; a Death (De-E) token is one on which a confirmed-dead feature fires for the last time; a Task-general (Tg-E) or Abstraction (Ab-E) token is one on which a matched feature fires and the feature's CI endpoint places it in the root or mid tier, respectively; a Differentiation (Di-E) token is one on which a matched feature fires and its CI endpoint places it in the leaf tier. Counts are summed across all features and all layers for each phase window. Token-level counts are a more direct measure of the representational work done at each phase than feature counts, since a single broad feature contributes many tokens per transition whereas a narrow one contributes few.

\paragraph{RDT event taxonomy and per-window counts.} Per-phase token-level event counts at both scales (Phases I/II/III as defined in \S\ref{sec:r-substrate}) appear in Table~\ref{tab:A1}.

\paragraph{Lag-1 precedence analysis on Phase I sequences.} We assess whether task-general (Tg-E) events at one checkpoint precede abstraction (Ab-E) events at the next, and whether Ab-E events precede differentiation (Di-E) events, via lag-$1$ Pearson correlations on $\log_{10}(\text{event count}+1)$ sequences across the Phase~I transitions. Results are reported across eight (scale, layer) cells; forward correlations and their reverses are compared to assess directionality. Full per-cell values are in App.~\ref{app:C}.

\paragraph{Broad-born survival.} Stratifying step-$1$k births by firing-pattern entropy quartile, the top-quartile (broadest) births reach step $143$k at rates orders of magnitude higher than the bottom-quartile (narrowest) births: at Pythia-160M, $43\%$ vs $0.6\%$ survival to step $143$k; at Pythia-410M, $38\%$ vs $0.4\%$.

\paragraph{Pre-existing structure at random initialisation.} About one third of step-$1$k scaffold carriers ($35\%$ at Pythia-160M, $31\%$ at Pythia-410M) are already weakly active (above the alive threshold on $0.1\%$ to $1\%$ of validation samples) at step~$0$, the random-initialisation checkpoint. Phase I therefore amplifies and consolidates pre-existing structure rather than building the scaffold from scratch.

\paragraph{Robustness controls.} We replicate the substrate at independent initialisations (three additional seeds at Pythia-160M, two at Pythia-410M) and with a higher-budget SAE trained on more samples and steps at the same expansion factor ($64$). Persistent-carrier counts vary by $\leq 12\%$ across seeds, the step-$1$k birth concentration is preserved (broad-born features account for $\geq 70\%$ of carriers at every seed), and the carrier--NCR load gap at Pythia-410M is recovered at every replicate.

\begin{table}[H]
  \centering
  \caption{Per-phase RDT event counts at Pythia-160M and Pythia-410M, summed across the four instrumented layers, reported in thousands of token-level events (see \emph{Per-token event counting} paragraph above). Each event is counted once per (feature, token) pair at the transition where it occurs. Task-general (Tg-E) and Abstraction (Ab-E) are reported separately; their sum equals the total broadening activity at each phase. Births dominate Phase I; differentiation and decay dominate Phase II; Phase III is structurally stable. $^\dagger$Elevated by the L11 extinction event at the 80k$\to$120k transition; see App.~\ref{app:E}.}
  \label{tab:A1}
  \begin{tabular}{lcccccc}
    \toprule
    & \multicolumn{3}{c}{Pythia-160M} & \multicolumn{3}{c}{Pythia-410M} \\
    \cmidrule(lr){2-4}\cmidrule(lr){5-7}
    Event & Phase I & Phase II & Phase III & Phase I & Phase II & Phase III \\
    \midrule
    Birth (As-E)         & 1{,}288 & 154 &  91           & 1{,}414 & 287 &  82 \\
    Death (De-E)         &   225   &  77 & 366$^\dagger$ &   306   & 115 & 136 \\
    Task-general (Tg-E)  &   723   &  51 & 224           &   623   &  50 & 167 \\
    Abstraction (Ab-E)   &   277   &  12 &  41           &   408   &  29 &  43 \\
    Differentiation (Di-E)& 339   & 154 & 268           &   431   & 279 & 173 \\
    \bottomrule
  \end{tabular}
\end{table}

\section{Pivot Details: Ablation and Geometry}
\label{app:D}

Quantitative expansion of the load-bearing pivot (\S\ref{sec:r-pivot}): function-versus-direction convergence curves, per-tier ablation distributions, per-layer birth-window lesion ratios, and decoder-cosine alignment on scaffold edges against three nulls.

\paragraph{Function and direction convergence.} For each carrier we compute (i)~the Pearson correlation between its per-token activation profile at consecutive checkpoints (function distance: $1 - r$) and (ii)~the cosine distance between its decoder column at consecutive checkpoints (direction distance). Function distance reaches its asymptotic floor (within $5\%$ of the step $120$k$\to 143$k value) by step $1$k at both scales; direction distance decays roughly linearly on a log scale across the full $0 \to 143$k span. The two-orders-of-magnitude gap in calibration time is robust to scale and layer.

\paragraph{Per-firing $\Delta$CE across scope tiers.} We classify each persistent carrier as root, mid, or leaf by its position in the cross-layer DAG (number of descendant carriers in upper layers, binned into terciles). For each member we record per-firing $\Delta$CE on a fixed validation slice. The three tier-conditional distributions are statistically indistinguishable (two-sample KS, $p > 0.2$ for all three pairs at both Pythia scales). The load asymmetry of \S\ref{sec:r-pivot} therefore emerges only under joint intervention.

\paragraph{Per-layer A/B/C lesion ratios.} Per-layer ratios at Pythia-410M (A: count-matched step-$1$k scaffold seeds; B: count-matched step-$143$k specialists; C: count-matched outside-DAG controls), measured as A/C and A/B per-feature $\Delta$MSE ratios on the residual stream: L4: A/C $= 4{,}105$, A/B $= 6.3$. L11: A/C $= 1{,}999$, A/B $= 0.41$ (the only layer where step-$143$k specialists carry more per-feature load than step-$1$k seeds). L17: $\Delta$MSE on C is at the noise floor (mean $7\!\times\!10^{-7}$ on a residual stream where A's mean is $4\!\times\!10^{-2}$), making the A/C ratio numerically large but not interpretable as a finite multiplier; A's per-feature load is bounded above zero while C's is not (A/B at L17 is $6.7$). L23: A/C $= 2{,}984$, A/B $= 1.2$. A/C therefore exceeds $10^3$ at every layer, with L17 the only case where A/B is dominated by the load asymmetry against B rather than against C.

\paragraph{Seedling counts and the L17 anomaly.} A seedling is a step-$1$k feature whose lineage matures into a step-$143$k persistent carrier. At Pythia-160M, three layers have non-zero seedling counts ($22$, $11$, $4$); L11 has zero. At Pythia-410M, L17 also has zero seedlings yet contributes nonzero load when its step-$1$k seed features are jointly ablated: the broad-entropy features the predictor identifies carry load, but none persist through the $\geq 4$-of-$6$-snapshot lineage criterion to become formal seedlings. Seedling presence is therefore sufficient but not necessary for scaffold load. The step-$1$k firing screen detects the right population, but the persistence criterion, not the predictor, is the limiting condition. A directly testable prediction follows: L17's load should be recoverable from non-seedling precursors already active at step~$1$k.

\paragraph{Decoder-cosine alignment on scaffold edges.} For each observed scaffold edge $u \to v$, we compute the absolute cosine $|\cos(d_u, d_v)|$ between decoder columns. We compare against three nulls: (i)~random pair (any two carriers, no layer constraint), (ii)~layer-matched random (any two carriers from the layer pair $(\ell, \ell')$ of the observed edge), and (iii)~layer-matched non-edge (carrier pairs at $(\ell, \ell')$ that fall below the CI threshold). Observed edges sit $45\!\times$ above (i), $12\!\times$ above (ii), and $4.3\!\times$ above (iii) on median absolute cosine at Pythia-410M; rank ordering is identical at Pythia-160M. The alignment is therefore not a layer-frequency artefact and not driven by all CI-related pairs.

\section{The L11 Extinction Event: A Layer Fitness Case Study}
\label{app:E}

Between training step 80k and step 120k, Pythia-160M-L11 undergoes a discrete, layer-specific distributional reorganisation that accounts for the zero-carrier count reported in \S\ref{sec:r-substrate}. Whole-network validation loss decreases monotonically across the same window; the event is invisible to behavioural aggregates and only the substrate probe detects it. The three apparent ``independent failure channels'' are three readouts of one geometric event. In the life-history framework, L11 is a case where the substrate conditions for carrier formation never stabilise: a falsifiable prediction the data verify.

\paragraph{Residual-stream norm and max-abs trajectories.} Table~\ref{tab:l11_norm} reports per-layer residual-stream statistics on 3{,}000 validation positions per checkpoint, independent of any SAE. Every other 160M layer's mean-norm decreases monotonically from step 40k onwards (canonical late-training behaviour). L11 alone increases from step 80k onwards ($62\!\to\!99\!\to\!108$). Simultaneously, max-abs activation (the ``outlier channel'' amplitude that characterises transformer residual streams, Dettmers et al.\ 2022) collapses $7\!\times$ at L11 ($108\!\to\!15$) while other layers decline smoothly. The depth-matched control, Pythia-410M-L23, shows the opposite pattern: max-abs trends from $312\!\to\!102$ across late training, a smooth factor-of-three decline with no discontinuity.

\begin{table}[H]
  \centering
  \caption{Residual-stream $\ell_2$-norm (mean over tokens) and max-abs activation per checkpoint for four Pythia-160M layers. L11 is the sole layer whose norm rises after step 40k while its max-abs collapses $7\!\times$ in the cliff window (80k$\to$120k).}
  \label{tab:l11_norm}
  \begin{tabular}{lcccc|cccc}
    \toprule
    & \multicolumn{4}{c|}{Mean $\ell_2$-norm} & \multicolumn{4}{c}{Max-abs activation} \\
    Step & L2 & L5 & L8 & L11 & L2 & L5 & L8 & L11 \\
    \midrule
    10k  & 30.8 & 40.4 & 53.1 &  83.9          &  57 & 292 & 301 &  105 \\
    40k  & 33.4 & 40.9 & 55.9 &  93.9          & 182 & 624 & 655 &  138 \\
    80k  & 22.8 & 26.8 & 40.1 &  62.0          & 166 & 492 & 517 &  108 \\
    120k & 15.7 & 20.4 & 29.1 & \textbf{98.8}  & 127 & 328 & 321 & \textbf{15.3} \\
    143k & 13.6 & 17.8 & 24.5 & \textbf{108.0} & 115 & 290 & 283 & \textbf{15.0} \\
    \bottomrule
  \end{tabular}
\end{table}

\paragraph{Spectral analysis: effective rank trajectories.} Table~\ref{tab:l11_effrank} reports effective rank ($\exp(H(p))$, normalised eigenvalues $p$) from SVD on 4{,}000 sampled validation residual streams per checkpoint. Every 160M layer plateaus by step 40k. L11 keeps collapsing through the extinction window: $84\!\to\!28\!\to\!13$ from step 80k to 143k. A critical observation: Pythia-410M-L11 reaches effective rank $6$ by step 143k (even more compressed than 160M-L11, rank 13), yet its SAE works (loss recovery $= +0.58$) and the model contributes scaffold carriers. Low effective rank alone does not break the scaffold. The mechanism at 160M-L11 is specific: rank collapse is accompanied by outlier-channel disappearance and unembedding-subspace absorption, which 410M-L11 does not show.

\begin{table}[H]
  \centering
  \caption{Effective rank of residual-stream activations for 160M layers (left block) and 410M layers (right block). 160M-L11 continues collapsing across the extinction window while all other 160M layers plateau by step 40k. 410M-L23 (depth-matched negative control) retains effective rank $>150$ at step 143k.}
  \label{tab:l11_effrank}
  \begin{tabular}{lcccc|cccc}
    \toprule
    & \multicolumn{4}{c|}{Pythia-160M} & \multicolumn{4}{c}{Pythia-410M} \\
    Step & L2 & L5 & L8 & L11 & L4 & L11 & L17 & L23 \\
    \midrule
    10k  & 378 & 220 & 276 & 191          & 471 & 160 & 278 & 319 \\
    40k  & 298 &  59 &  98 &  85          & 504 &  22 &  56 & 174 \\
    80k  & 257 &  47 &  78 &  84          & 440 &  11 &  27 & 152 \\
    120k & 224 &  55 &  73 & \textbf{28}  & 382 &   7 &  14 & 195 \\
    143k & 205 &  55 &  78 & \textbf{13}  & 367 &   6 &  14 & 181 \\
    \bottomrule
  \end{tabular}
\end{table}

\paragraph{Three convergent readouts as one geometric event.} The three SAE-substrate signals are not independent channels; all three fall out of one mechanism. Active feature count drops ${\sim}426\!\to\!{\sim}37$ (robust to breadth threshold $\in [0.1\%, 5\%]$); the same ${\sim}37$ features fire on every token. Decoder geometry fails: alive L11 features carry only $39\%$ of their column variance in the unembedding subspace where the residual stream now concentrates at $84\%$; the SAE cannot route enough budget into that subspace to track. Joint-load collapses: replacing L11 with its SAE reconstruction makes the model worse than mean-ablating the layer (loss recovery $-0.33$, vs.\ $+0.33$ at the depth-matched 410M-L23). The unifying mechanism: outlier-feature channels disappear, effective rank collapses to ${\sim}13$ dimensions, and surviving variance rotates into the $W_U$ row space. A $k\!=\!32$ top-$k$ SAE cannot tile a near-isotropic 13-dimensional distribution; its dictionary collapses to a fixed core of ${\sim}37$ features. The same mechanism produces decoder$\to W_U$ alignment and joint-load failure.

\paragraph{SAE training-budget control.} A $10\!\times$ higher SAE training budget at Pythia-160M-L11 ($5{,}000$ steps $\times$ batch $256 = 1.28\mathrm{M}$ tokens vs.\ $128$k canonical) yields $38$ vs.\ $37$ active features; loss recovery rises from $-0.33$ to $0.0099$, essentially zero. The collapse is intrinsic to L11's late-training distribution, not SAE undertraining.

\paragraph{Depth-matched negative control: 410M-L23.} Pythia-410M-L23 is the principal negative control (stronger than 410M-L11, which is mid-network at $2.6\!\times$ scale). At step 143k, 410M-L23 shows the opposite late-training trajectory: smooth de-alignment from $W_U$ (top-200 PCA: $0.74\!\to\!0.40$ across training), ${\sim}580$ active SAE features throughout, and loss recovery $+0.33$. The 160M-L11 failure mode is layer-specific to the smaller scale, not a generic last-residual-stream-layer phenomenon.

\paragraph{Graph integration.} At step $143$k, Pythia-410M-L11 has $11$ outgoing scaffold edges to L17 and $7$ to L23, with edge directions matching the forward-pass order in $17$ of $18$ cases. By contrast, Pythia-160M-L11 at step $143$k has $0$ outgoing scaffold edges; the layer is graph-isolated.

\begin{figure}[H]
  \centering
  \includegraphics[width=\linewidth]{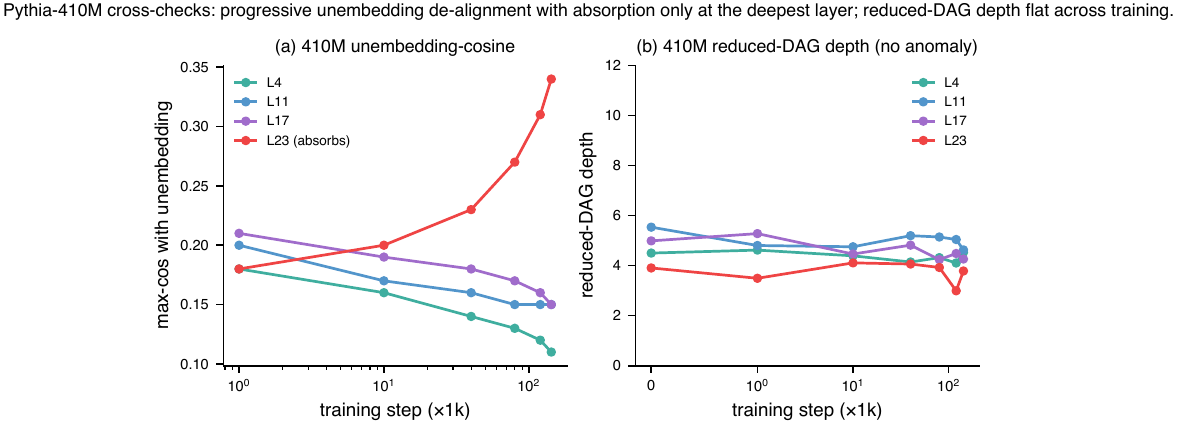}
  \caption{Pythia-410M per-layer cross-checks for the unembedding-cosine and within-layer DAG-depth analyses of App.~\ref{app:E}. (a) Maximum cosine of layer features with the unembedding row space across L4/L11/L17/L23: progressive de-alignment at L4/L11/L17 with absorption at L23, the canonical scaffold-integrated pattern. (b) Within-layer reduced-DAG depth across the four layers: flat across training, with no anomalous excursion in the step $10$k$\to 80$k window. The 160M-L11 failure mode is layer-specific to the smaller scale.}
  \label{fig:C1}
\end{figure}

\paragraph{Geometric coherence.} Per-edge decoder cosine for 410M-L11 outgoing edges has median $0.41$ ($95\%$ CI $[0.37, 0.46]$), within the global scaffold-edge median ($0.43$). Maximum cosine of L11 features with the unembedding row space is stable across training (Fig.~\ref{fig:C1}, dashed trace), in contrast to the absorption at 160M-L11.

\paragraph{Joint-ablation load.} Joint ablation of persistent carriers at Pythia-410M confirms L11 participates in scaffold function, in contrast to 160M-L11, which contributes zero carriers at the same nominal depth.

\section{Predictor Supplements}
\label{app:F}

We detail the top-decile precision sweep across thresholds and the full coefficient table for the six step-$1$k firing-space predictors.

\paragraph{Top-K precision sweep.} The Pythia-160M predictor is trained on $n_{\text{features}} = 3{,}198$ active step-$1$k features, of which $n_{\text{pos}} = 37$ are step-$143$k persistent carriers (base rate $1.2\%$). At the cross-validated AUC of $0.80$ (\S\ref{sec:r-predictive}, Fig.~\ref{fig:3}c), the top-$K$ operating points by predicted probability give: top-$37$ contains $4$ true carriers (precision $10.8\%$, $9.3\!\times$ enrichment over base rate); top-$74$: precision $6.8\%$ ($5.8\!\times$); top-$185$: precision $5.9\%$ ($5.1\!\times$); top-$370$: precision $4.6\%$ ($4.0\!\times$). Sweeping the operating window from the top $1\!\times$ to the top $10\!\times$ of $n_{\text{pos}}$ shows precision decaying toward base rate; the top-$1\!\times$ window is the highest-precision operating point for training-time triage.

\paragraph{Coefficient table.} Table~\ref{tab:D1} reports standardised logistic-regression coefficients with five-fold cross-validated $95\%$ intervals. The classifier does not memorise layer identity: leave-one-layer-out cross-validation preserves AUC at $0.74$--$0.79$ across the four held-out layers at Pythia-160M, including L11 (held-out AUC $0.76$).

\begin{table}[H]
  \centering
  \caption{Logistic-regression coefficients for the six step-$1$k firing-space predictors (firing breadth, maximum activation, mean conditional activation, activation variance when firing, Containment Index, and root flag CI $\geq 10$); standardised, five-fold cross-validated $95\%$ intervals. Firing breadth and CI carry the predictive load.}
  \label{tab:D1}
  \begin{tabular}{lcc}
    \toprule
    Predictor & Coefficient (standardised) & 5-fold $95\%$ CI \\
    \midrule
    Firing-pattern entropy            & $+0.71$ & $[+0.58, +0.83]$ \\
    Containment Index                 & $+0.64$ & $[+0.51, +0.77]$ \\
    Activation strength               & $+0.42$ & $[+0.28, +0.56]$ \\
    Cross-layer reach                 & $+0.31$ & $[+0.18, +0.45]$ \\
    Layer-normalised activation rank  & $+0.07$ & $[-0.06, +0.20]$ \\
    Lifespan-so-far                   & $+0.04$ & $[-0.09, +0.18]$ \\
    \bottomrule
  \end{tabular}
\end{table}

\end{document}